\documentclass[sigconf]{acmart}
\usepackage{array}
\usepackage{graphicx}
\usepackage{clrscode}
\usepackage{balance}
\usepackage{subcaption}
\usepackage{multirow}
\usepackage{booktabs}
\usepackage{adjustbox}
\usepackage[justification=justified,skip=3pt]{caption}
\usepackage{float}
\usepackage{color}
\usepackage{soul}
\usepackage{mdframed}
\usepackage{amsopn}
\usepackage{mathrsfs}
\usepackage{mathtools}
\usepackage{amsmath}
\usepackage{arydshln}
\usepackage{hyperref}
\usepackage{multicol}
\usepackage{blkarray}
\usepackage{enumerate}
\usepackage{courier}
\usepackage{rotating}
\usepackage{booktabs}
\usepackage{diagbox}
\usepackage{fancybox}
\usepackage{minibox}
\usepackage{cases}
\usepackage[lined,boxruled,commentsnumbered,linesnumbered]{algorithm2e}

\copyrightyear{2020}
\acmYear{2020}
\setcopyright{acmlicensed}\acmConference[SIGIR '20]{Proceedings of the 43rd
International ACM SIGIR Conference on Research and Development in
Information Retrieval}{July 25--30, 2020}{Virtual Event, China}
\acmBooktitle{Proceedings of the 43rd International ACM SIGIR Conference on
Research and Development in Information Retrieval (SIGIR '20), July 25--30,
2020, Virtual Event, China}
\acmPrice{15.00}
\acmDOI{10.1145/3397271.3401056}
\acmISBN{978-1-4503-8016-4/20/07}

\begin{document}
\fancyhead{}
\title{Learning Personalized Risk Preferences for Recommendation}

\author{Yingqiang Ge$^{\dagger}$, Shuyuan Xu$^{\dagger}$, Shuchang Liu${^\dagger}$, Zuohui Fu${^\dagger}$, Fei Sun${^\ddagger}$, Yongfeng Zhang${^\dagger}$}
\affiliation{%
  \institution{$^{\dagger}$Rutgers, The State University of New Jersey \qquad $^{\ddagger}$Alibaba Group}
}
\email{{yingqiang.ge, shuyuan.xu, sl1471, zuohui.fu}@rutgers.edu, ofey.sunfei@gmail.com, yongfeng.zhang@rutgers.edu}







\begin{abstract}
The rapid growth of e-commerce has made people accustomed to shopping online.
Before making purchases on e-commerce websites, most consumers tend to rely on rating scores and review information to make purchase decisions.
With this information, they can infer the quality of products to reduce the risk of purchase.
Specifically, items with high rating scores and good reviews tend to be less risky, while items with low rating scores and bad reviews might be risky to purchase.
On the other hand, the purchase behaviors will also be influenced by consumers' tolerance of risks, known as the risk attitudes.
Economists have studied risk attitudes for decades. 
These studies reveal that people are not always rational enough when making decisions, and their risk attitudes may vary in different circumstances. 

Most existing works over recommendation systems do not consider users' risk attitudes in modeling, which may lead to inappropriate recommendations to users. 
For example, suggesting a risky item to a risk-averse person or a conservative item to a risk-seeking person may result in the reduction of user experience.
In this paper, we propose a novel risk-aware recommendation framework that integrates machine learning and behavioral economics to uncover the risk mechanism behind users' purchasing behaviors. 
Concretely, we first develop statistical methods to estimate the risk distribution of each item and then draw the Nobel-award winning Prospect Theory into our model to learn how users choose from probabilistic alternatives that involve risks, where the probabilities of the outcomes are uncertain. 
Experiments on several e-commerce datasets demonstrate that by taking user risk preferences into consideration, our approach can achieve better performance than many classical recommendation approaches, and further analyses also verify the advantages of risk-aware recommendation beyond accuracy. 
\end{abstract}



\maketitle

\section{Introduction}
Designing personalized recommender systems is able to help users find relevant items efficiently in the context of web information overload. 
A well-informed recommender system is capable of not only saving consumers' exploration time but also benefiting the revenue of various online economic platforms. 
Traditional recommendation algorithms mostly focus on optimizing rating- or ranking-oriented metrics. 
However, previous recommendation algorithms seldom consider users' economic incentives when modeling user behaviors and generating recommendations.

In fact, consumers' economic incentives play an important role when making decisions in online economic systems such as e-commerce \cite{zhang2016economic, ge2019maximizing}. 
 A common observation in practical recommendation systems is that customers may use rating scores and review information to support their buying decisions \cite{mo2015effect,lackermair2013importance,Leino:2007:CAR:1297231.1297255}.
According to this information, buyers can infer the quality of products to avoid wasting time and reduce the risk of purchase.
However, the judgments of risky items are different from person to person.
Actually, users with different risk attitudes could make different decisions, even when facing the same situation. 
Concretely, risk-averse consumers would be more likely to buy products that are ``safe choices'', i.e., that have many good reviews, while risk-appetite consumers may more likely try new products even though they yet to have sufficiently many good reviews. 
Meanwhile, based on real-user psychological experiments, behavioral economists \cite{kahneman1979prospect} have revealed that most consumers will overweight small probability gains and underweight moderate or high probability losses. 
For instance, when facing 1\% probability to win \$500 or 100\% probability to win \$5, most people would take the risk and choose the first choice, although the expected gains are the same; however, when facing 1\% probability to lose \$500 or 100\% probability to lose \$5, most people would mitigate the risk and choose the second one, although the expected losses are the same. 
Thus, understanding consumer decisions under risk can assist researchers to better build personalized recommendation algorithms according to users' risk preferences, and an informed recommendation system with risk-awareness can provide appropriate recommendations that make consumers comfortable. 
Fortunately, integrating machine learning and the established economic principles can help to model the risk attitude of the user decision process based on large-scale user transaction logs.

In this paper, we propose a novel risk-aware recommendation framework, which takes users' risk attitudes into consideration according to several behavioral economic principles. 
In particular, we simulate the risk distribution of each item based on its rating distribution by conventional statistical methods. 
Combining the Nobel-prize winning theory in economics --- the Prospect Theory \cite{kahneman1979prospect} --- with machine learning algorithms, our model learns to predict the consumer decisions with risk-awareness. 

The key contributions of the paper can be summarized as follows:
\begin{itemize}
\item We take consumers' risk attitudes and weighted event probability into consideration for economic recommendation, which better simulates real-world online economic environment, where users have to make decisions under potential risks of dissatisfaction. 
\item Our model can be considered as a personalized version of the Prospect Theory, i.e., unlike the original Prospect Theory, which assumes a uniform risk function with the same parameters for all subjects, our method endeavors to figure out the personalized risk preference for each user, which adapts the economic theory under different risk attitudes for different users.
\item Experimental results on real-world e-commerce verify that our approach can not only achieve better recommendation performance than both classical and economic recommendation baselines but also successfully adapt the Prospect Theory to learn each user's personalized risk attitudes.
\end{itemize}

The following contents of the paper will be organized as follows: we first review some related work in section \ref{sec:related}, and then introduce the key economic preliminaries in section \ref{sec:preliminary} to prepare readers with the underlying economic backgrounds used in this work. 
The proposed model and recommendation strategies are introduced in section \ref{sec:framework}, followed by experimental results in section \ref{sec:experiments}. 
Finally, we conclude the work with possible future research directions in section \ref{sec:conclusions}.

\section{Related Work}\label{sec:related}
In this section, we will briefly introduce some background knowledge to help the readers get a better understanding of the areas that are related to our work.

\subsection{\mbox{Collaborative Filtering}}
Collaborative Filtering (CF) has been one of the most dominant approaches to recommender systems. 
Early methods of CF consider the user-item rating matrix and conduct rating prediction task with user-based \cite{resnick1994grouplens,konstan1997grouplens} or item-based \cite{sarwar2001item,Linden2003} collaborative filtering methods. 
In these methods, the user and item rating vector are considered as the representation vector for each user and item.

With the development of dimension reduction methods, latent factor models, such as singular value decomposition (SVD) \cite{koren2009matrix}, non-negative matrix factorization \cite{lee2001algorithms}, and probabilistic matrix factorization \cite{mnih2008probabilistic}, are later widely adopted in recommender systems. 
In these aforementioned matrix factorization approaches, each user and item is learned as a latent factor representation to calculate the matching score of the user-item pair.

Deep models have recently been further extended to collaborative filtering methods for the recommendation tasks. 
The relevant methods can be roughly divided into two subcategories: similarity learning methods and representational learning methods.
The similarity learning approaches adopt simple user/item representations (such as one-hot) and learn a complex prediction network as the similarity function to calculate user-item matching scores \cite{he2017neural}.
Meanwhile, the representation learning approaches learn rich user/item representations and adopt a simple similarity function (e.g., inner product) for matching score calculation \cite{zhang2017joint}.

Another important research direction is learning to rank (LTR) for recommendation, which learns the relative ordering of items rather than the absolute preference scores. 
Probably, the most representative method in this direction is Bayesian Personalized Ranking (BPR) \cite{bpr}, which is a pair-wise learning-to-rank method for recommendation. 
It is also further generalized to take other information sources such as images \cite{he2016vbpr} and text \cite{zhang2017joint, FuXianGaoZhaoHuangGeXuGengShahZhangDeMelo2020FairRecommendation} into consideration.

\subsection{Economic Recommendation}
For a long time, the focus of recommendation system research has been on the above-mentioned rating- or ranking-related tasks, such as rating prediction and top-N recommendation, however, the related approaches rarely consider the economic value that the recommendation lists could bring to the users or the system, although it is one of the most important goals for real-world recommendation systems.
Fortunately, we can find some recent research on economic recommendation has begun to take care of the economic value of personalized recommendation. 
For example, \cite{wang2011utilizing} studied the user's sense of value for items in terms of utility in recommender systems, and \cite{zhao2015commerce} conducted large-scale experiments with real-world users to verify consumers' sense of utility for personalized promotions. 
Further, \cite{zhang2016economic} bridged economic principles and machine learning methods to maximize the social surplus for recommendation, and \cite{zhao2017multi} proposed to learn the substitute and complementary relations between products so as to maximize consumer utility. 
Researchers have also considered using economic models for user behavior analysis in search systems \cite{azzopardi2014modelling}.
Although current economic recommendation approaches may improve the economic value, their underlying motivation is to maximize a total utility function for each user to generate recommendations, which is quite different from ours. 
Established behavioral economic principles show that consumers usually make decisions under risk conditions, and the risk attitude/preference of different users could be different, which may influence their decisions in economic systems \cite{kahneman1979prospect}. This motivates us to estimate the risk preference of users for economics-driven risk-aware recommendation.

There are only a few works studying the effect of user risk attitudes in recommendation systems. 
\cite{DBLP:journals/corr/abs-1812-11422} is motivated by the intuition that ``the pain of loss is more powerful than the pleasure from a similar gain'', and designs an algorithm suggesting fewer negative items to users.
\cite{DBLP:journals/corr/abs-1904-05325} shows that most of the existing recommendation algorithms are either risk-neutral or risk-averse, and it designs a risk-seeking strategy to recommend items. 
However, the users' risk attitudes are personalized, which can hardly be modeled by a single global assumption. 
As a result, we decide user risk attitudes according to different scenarios (gains and losses) based on Prospect Theory, take advantage of machine learning to personalize their risk attitudes, and use prospect values to make top-K recommendations.
Another stream of risk-aware recommendation research belongs to Bouneffouf et al. \cite{10.1007/978-3-642-42054-2_8,DBLP:journals/corr/Bouneffouf14b}. In this research, the risk is defined as the exploration-exploitation trade-off in contextual bandit algorithms, which is different from our perception of consumers' risk attitudes and the consumer decision-making process under uncertainty.

\subsection{Decision-Making Under Uncertainty}
The real-world is filled with plenty of uncertainties, and it involves various risk factors. 
People have to make a large number of decisions under uncertainty in their daily life, such as making purchases or investments. 
As a result, decision-making under risk is one of the most fundamental research subjective for economists. 
There have been two major types of theory on decision-making under risk, and the critical difference between them is how to understand the definition of uncertainty. 
The first type of theory holds the view that uncertainty is objective, which means that the uncertainty of an event can be represented as an objective probability distribution of all possible outcomes. 
One representative theory of this view is the von Neumann-Morgenstern Expected Utility Theory \cite{von2007theory}. 
The other type of risk theory believes that the uncertainty of events is not objective, but it is based on people's subjective judgments of all possible outcomes. 
In particular, Savage \cite{savage1972foundations} puts forward a theory of subjectivity based on personal probability and statistics, which constitutes the research line underlying Bayesian statistics. 

Overall, there exists more than one economic theory on risk modeling, and among them, the Expected Utility Theory was generally adopted and practically applied due to its intuition and convenience. 
However, economists find that these theories still cannot correctly explain all the known consumer behaviors, and most importantly, researchers gradually realized the limitations of these theories, because of the emerging of several classical paradoxes out of the theories, which were raised by Bernoulli \cite{bernoulli1968specimen}, Allais \cite{allais1953extension}, and Ellsberg \cite{ellsberg1961risk}. 
To solve these paradoxes, Daniel Kahneman and Amos Tversky renewed the study based on human behaviors and proposed the Prospect Theory \cite{kahneman1979prospect} in the year of 1979 to model the user risk preferences, and later in 2002, Kahneman was awarded the Nobel Prize in Economics for the success of Prospect Theory in modeling consumer risks in economic behaviors.


\section{Preliminaries}\label{sec:preliminary}
In this section, we introduce some essential economic preliminaries about the above-mentioned risk attitudes, and prospect theory to help readers better understand our proposed model later.

\subsection{Risk Attitudes}
Users usually have to take the risk for their own choices when making decisions under uncertainty. In recommender systems, consumers hardly know for sure if they would be satisfied with a product or not, because a purchase that is believed to be a good match may still result in dissatisfaction due to various factors, such as the quality or delayed shipping, which can be considered as potential risk factors in e-commerce. As we have discussed before, different people may have different attitudes toward risks \cite{virine2016projectthink,hillson2017understanding,adhikari2016does}. In general, consumer risk attitudes can be divided into three categories, which are introduced as follows: 

\textbf{Risk Aversion}: Even though there exist many methods to describe an individual's risk aversion attitude, the essence of them is the same, i.e., comparing all of the possible returns under different choices that have the same expected value, a risk-aversion consumer would always prefer the certain returns. 

\textbf{Risk Appetite}: The definition of risk appetite is counterpart against risk aversion, i.e., comparing all of the possible returns under different choices that have the same expected value, a risk-appetite consumer would always prefer the risky returns. 

\textbf{Risk Neutral}: Risk neutral is a mindset where an individual is indifferent to risk when making a decision under uncertainty.

It would be easier to understand the difference between different risk preferences with an example. We still take gambling as an instance, but this time, we focus on people's risk attitudes instead of overweight small probability gains and underweight moderate or high probability losses. Suppose there are two choices in a gamble, one is to lose either \$100 or \$0 with 50\% probability each, and the other choice is to lose \$50 with 100\% probability. Though the expected utility of both choices is the same (\$-50), different users may make different choices based on their risk attitude. A risk-aversion user would prefer to choose a sure \$50 lose, and a risk-appetite user would prefer to risk for a potential \$0 lose, while a risk-neutral user is indifferent to the two choices.

\subsection{Prospect Theory}
Prospect theory is a commonly accepted and widely applied economic principle in behavioral economics, which describes the consumer's decision-making process between probabilistic alternatives involving risks. 
It believes that people have different evaluations for gains and losses, and describes the decision-making process in two steps.
Firstly, people will divide all potential outcomes into gains or losses by setting a reference point that they consider as indifferent.
Consequently, those outcomes less than the reference point are considered as losses, and those that are greater are considered as gains.
Secondly, people have different prospect value functions for each outcomes and its probabilistic estimation. 
With this, people will calculate prospect values based on all the possible outcomes and their respective probabilities, and finally, choose the best action according to prospect values instead of raw outcomes.

According to Kahneman \cite{kahneman1979prospect,tversky1992advances}, the prospect value function is a product of the value function and the weighting function.
While the value function measures the subjective value of the gain/loss,
the weighting function can be seen as a subjective distortion of the original probabilities, which will be introduced later. 

\begin{figure}[h]
\centering
\hspace{-10pt}
\centering
\begin{subfigure}{0.25\textwidth}
\includegraphics[scale=0.43]{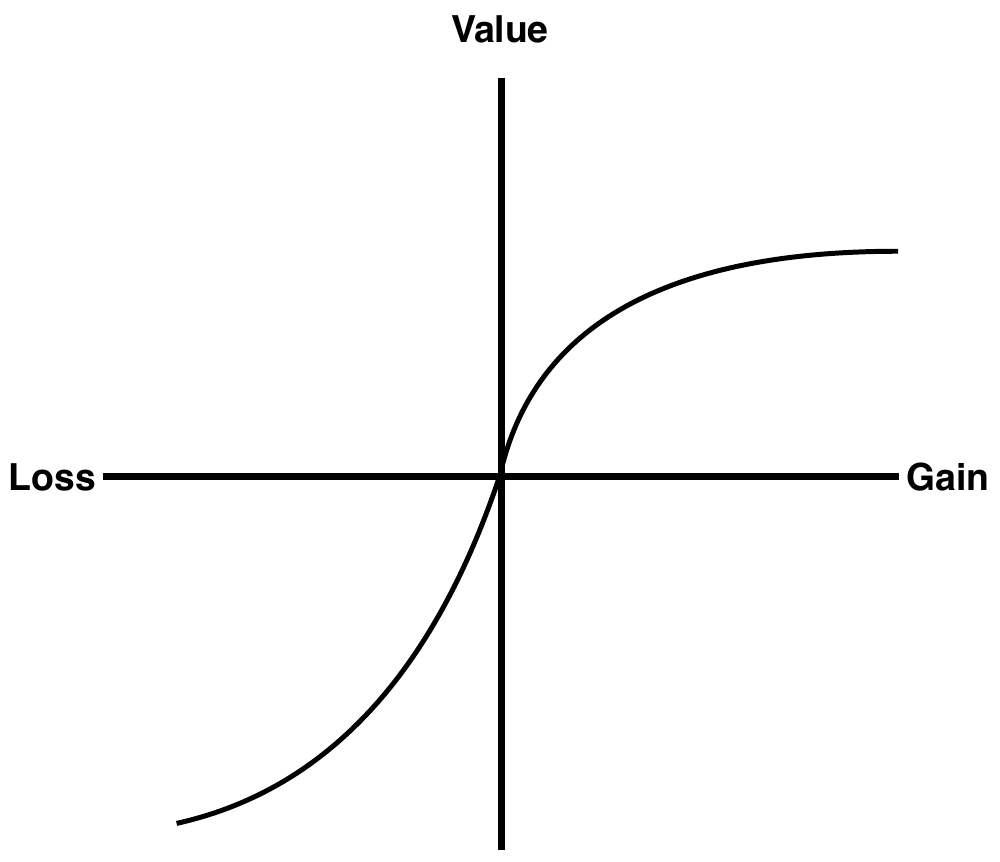}
\subcaption{value function}
\label{fig:value1}
\end{subfigure}
\hspace{-8pt}
\begin{subfigure}{0.25\textwidth}
\includegraphics[scale=0.49]{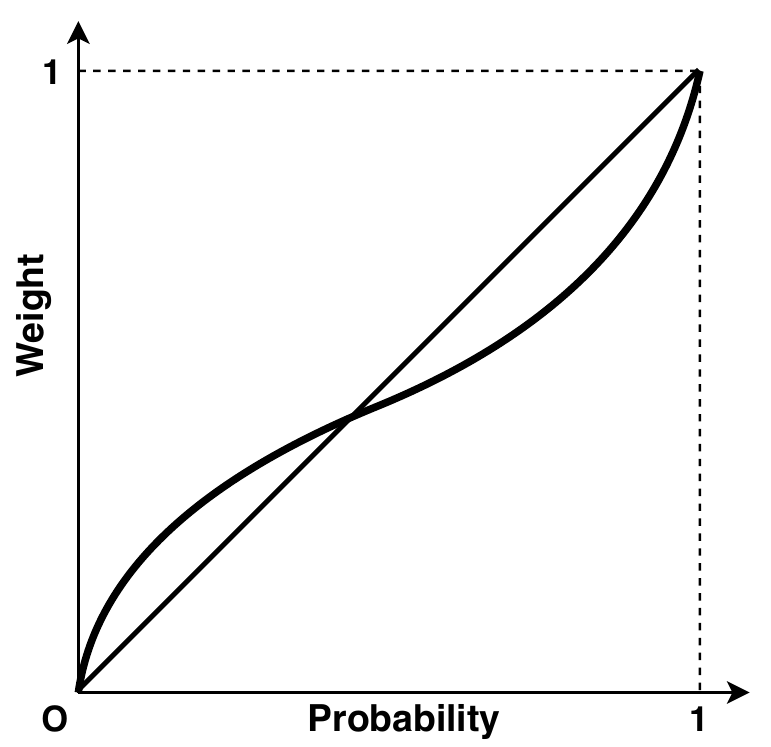}
\subcaption{weighting function}
\label{fig:weight1}
\end{subfigure}
\caption{The left figure shows a hypothetical value function, while the right figure is a hypothetical weighting function.}
\label{fig:originalPT}
\vspace{-10pt}
\end{figure}

On one hand, through a large number of psychological experiments, \citeauthor{kahneman1979prospect} depicted some general consumer psychological principles and summarized that people usually are risk-aversion for gains and risk-appetite for losses, because they have stronger feelings for losses than for gains, i.e., the happiness out of gaining \$100 usually cannot compensate for the hurt of losing \$100. 
The hypothesis of the value function is based on the above observations and is shown in Fig. \ref{fig:value1}. 
For gains, the function is concave to show risk-aversion; for loss, it is convex to show risk-appetite.
Also, it is steeper for losses than gains indicating that losses outweigh gains.

On the other hand, according to the observations of many behavioral economic experiments \cite{cohen1987experimental, 10.2307/1880632}, people usually overweight the low probabilities and underweight the high probabilities for both gains and losses.
This trend, which is not reflected in the value function, is reflected by the individual differences in the weighting function. 
Specifically, there exists a probability distortion that people generally do not look at the value of probability uniformly between 0 and 1. 
Lower probability will be overweighted (that means a person is over concerned with the low probability outcome),
while higher probability is underweighted (that means a person is not concerned enough with the high probability outcome). 
Based on these observations, the hypothesis of the shape of the weighting function is shown in Fig.\ref{fig:weight1}. 
\section{The Framework}\label{sec:framework}
In this section, we further discuss how to bridge the fundamental economic concepts with machine learning algorithms to establish a risk-aware recommendation framework.

\subsection{Personalized Prospect Theory}
In order to implement prospect theory in real-world recommendation scenario, we first define a few concepts and then specify their function in e-commerce.
Let $S$ be a finite set of states of nature; subsets of $S$ are called events. 
In \cite{tversky1992advances}, it is assumed that exactly one state obtains, which is unknown to the decision-maker (e.g., one item only can be rated with 1 or 2 or ... or 5 stars on Amazon by a certain user).
Let $X$ be a set of consequences/outcomes and assume that $X$ includes a neural outcome, denoted as $0$, then we interpret all other elements as gains or losses through comparisons.
An uncertain prospect $f$, which is a function from $S$ to $X$, is represented as a sequence of pairs $(x_{i},s_{i})$, which yields $x_{i}$ if $s_{i}$ occurs, and $(s_{i})$ is a partition of $S$.
When a prospect $f=(x_{i},s_{i})$ is given by a probability distribution $p(s_{i})=p_{i}$, it can be viewed as a risky prospect $(x_{i},p_{i})$.

In an e-commerce scenario, the consumer satisfaction of purchasing an item is usually reflected by the corresponding rating scores \cite{Flanagin2014}, which can be seen as a potential state of the purchase. 
Therefore, we define the ratings, which are usually in the range of $\{1,2,3,4,5\}$, as states and get $S = \{r_{1},r_{2},r_{3},r_{4},r_{5}\}$, where $r_{i}=i$.
For the potential outcome of one purchase, we assume it is related to both the price of a given item and the expectation of the consumer to that item.
We use the reference point, denoted by $\hat{r}_{u}$, to represent the user $u$'s expectation to a given item.
The difference between the final rating and the reference point will cause a gain or a loss to the consumer.
Then, for every possible state $r_{i}$, the outcomes (gains or losses) can be designed as:
\begin{equation}\label{eq:value}
\begin{aligned}
    &x_{i} = price\cdot\tanh(r_{i}-\hat{r}_{u}),\\
    &\text{where} \quad r_{i} = i, i \in {1,\cdots,5}
\end{aligned}
\end{equation}

The outcome $x_{i}$ is positive for gains (i.e., the rating is greater than the reference point) and negative for losses (i.e., the rating is less than the reference point).
Since the difference between $r_{i}$ and $\hat{r}_{u}$ won't be too large, we choose the hyperbolic tangent function $\tanh(\cdot)$ to normalize $(r_{i}-\hat{r})$ into $[-1,1]$ so that we get $x_{i} \in [-price, price]$.
The probability distribution $p(r_{i})$ under the e-commerce setting  will be discussed in detail in Sec. \ref{sec:riskDistribution}; for now, we assume there exists certain $p(r_{i}) = p_{i}$, where $r_{i} \in S$.


\subsubsection{The Value Function}
To accord with the basic ideas of the Prospect Theory --- a value
function is concave for gains, convex for losses, and steeper for losses than for gains, we use the following function $v(x)$ as the value function, which comes from Kahneman's work in 1992 \cite{tversky1992advances}.
\begin{equation}\label{eq:value}
    v(x_{i})=
    \begin{cases}
    \lambda x_{i}^\alpha & \text{if } x_{i}\ge0  \qquad (v_{+})\\
    -(-x_{i})^\beta & \text{if } x_{i}<0  \qquad (v_{-})\\
    \end{cases}\\
\end{equation}
where $0< \alpha, \beta, \lambda < 1$. $\lambda \in (0,1)$ is to make sure that the value function is steeper for losses than for gains. 
Since our model needs to learn the personalized risk attitudes of a user $u$ under the circumstance that he/she is given a certain item $v$, we will use $\alpha_{uv}$, $\beta_{uv}$, $\lambda_{uv}$ in the following sections to represent their use in certain user-item pair.  
The above function reflects the required value function shape in prospect theory, as shown in Fig.\ref{fig:value1}.


\subsubsection{The Weighting Function}
The weighting function is used to represent the decision weight for all possible outcomes based on their corresponding probabilities, and we use the following function $\pi(p)$ as the weighting function, which is still consistent in the way of Kahneman's study \cite{tversky1992advances}.
\begin{equation}\label{eq:weight}
    \pi(p_{i})=
    \begin{dcases}
    \frac{p_{i}^\gamma}{(p_{i}^\gamma+(1-p_{i})^\gamma)^{1/\gamma}}& x_{i}\ge0  \qquad (\pi_{+})\\
    \frac{p_{i}^\delta}{(p_{i}^\delta+(1-p_{i})^\delta)^{1/\delta}}& x_{i}<0  \qquad (\pi_{-})
    \end{dcases}\\
\end{equation}
where $p_{i}$ is the probability of the gain (or loss), depending on the corresponding $x_{i}$; $\pi(0)=0$ and $\pi(1)=1$; $\gamma$ and $\delta$ are positive parameters ($0< \gamma, \delta < 1$) used to control the shape of the weighting function, which will be learned in the model. It should be noted that the weight $\pi(p)$ is no longer a probability, and it only represents the importance of a gain/loss for the consumer.
For the same purpose of personalization, we will use $\gamma_{uv}$ and $\delta_{uv}$ to replace them in the following sections.
 
\subsubsection{Prospect Value}
According to Kahneman and Tversky's work in 1979 \cite{kahneman1979prospect} and 1992 \cite{tversky1992advances}, we use a simple mathematical formula to calculate the prospect value defined as:
\begin{equation}\label{eq:prospectfun}
    \mathcal{V}=\sum_{i=1}^l v(x_i)\pi(p_i)
\end{equation}
where $\mathcal{V}$ represents the prospect value for a certain choice, which consists of $v(x)$ as the value function and $\pi(p)$ as the weighting function; $v(x_1),\cdots,v(x_l)$ are the values of gains or losses for all potential states, and $p_1,\cdots,p_l$ are the corresponding probabilities.

\subsubsection{Calculation Paradigm}
In order to clarify the calculation of the proposed prospect value and help readers better understand our framework, we give an example to show the process.
Suppose we have a user-item pair, and assume we know the user's expected score of that item is 2.5 (which indicates $\hat{r}_{u}=2.5$). Suppose we also know the rating distribution of that item, e.g. \{1 star : $p_1$, 2 stars : $p_2$, ..., 5 stars : $p_5$\}. 
Then, we have Table \ref{table:pv_sample} showing all possible states with their corresponding outcomes and weights.
We can see that since 1 star and 2 stars are below 2.5, which make $x_1$ and $x_2$ below to zero, they are losses for the user and need to use $-(-x_{i})^\beta $ as the value function and $\frac{p_i^\delta}{(p_i^\delta+(1-p_i)^\delta)^{1/\delta}}$ as the weighting function; on the contrary, 3 to 5 stars are gains for the user, thus, we use the remaining parts.
Based on Eq. \ref{eq:prospectfun}, the prospect value of this user-item pair is calculated as follows:
\begin{equation}
\begin{aligned}
    \mathcal{V} = &v_{-}(x_{1})\pi_{-}(p_{1})+
                v_{-}(x_{2})\pi_{-}(p_{2})+
                v_{+}(x_{3})\pi_{+}(p_{3})\\
                &+v_{+}(x_{4})\pi_{+}(p_{4})+
                v_{+}(x_{5})\pi_{+}(p_{5})
\end{aligned}
\end{equation}
where $v_{-}(x_{i})=-(-x_{i})^\beta$ and $\pi_{-}(p_i)=\frac{p_i^\delta}{(p_i^\delta+(1-p_i)^\delta)^{1/\delta}}$ represent the negative part of the value function and the weighting function, respectively (when $x_{i} \textless 0$), while $v_{+}(x_{i})=\lambda x_{i}^\alpha$ and $\pi_{+}(p_i)=\frac{p_i^\gamma}{(p_i^\gamma+(1-p_i)^\gamma)^{1/\gamma}}$ represent the positive part when $x_{i} \geq 0$. 

\begin{table}[h]
\vspace{0pt}
\caption{\label{font-table} Examples showing how to calculate the prospect value for a certain user-item pair. 
}
\begin{center}
\renewcommand{\multirowsetup}{\centering}
\setlength{\tabcolsep}{7pt}
\begin{tabular}{m{1.2cm}ccc }
\toprule
 States & $x_i$ & Value & Weight \\ \midrule
 1 star  &$price\cdot \tanh(1-2.5) \textless 0$ &$v_{-}(x_{1})$&$\pi_{-}(p_{1})$ \\ 
 2 stars &$price\cdot \tanh(2-2.5) \textless 0$ &$v_{-}(x_{2})$&$\pi_{-}(p_{2})$ \\
 3 stars &$price\cdot \tanh(3-2.5) \textgreater 0$ &$v_{+}(x_{3})$&$\pi_{+}(p_{3})$ \\
 4 stars &$price\cdot \tanh(4-2.5) \textgreater 0$ &$v_{+}(x_{4})$&$\pi_{+}(p_{4})$ \\
 5 stars &$price\cdot \tanh(5-2.5) \textgreater 0$ &$v_{+}(x_{5})$&$\pi_{+}(p_{5})$ \\ \bottomrule
\end{tabular} 
\end{center}

\label{table:pv_sample}
\vspace{-10pt}
\end{table}

\subsection{Estimation of Rating Distribution}\label{sec:riskDistribution}
Our objective is to solve the classical problem of decision-making under uncertainty, and therefore, in this section, we define the uncertainty in e-commerce.
Consumers often use other information about the products, such as rating scores and reviews, to estimate their probability of satisfaction and support their purchase decision.
In this work, we assume buyers use rating scores to estimate the product quality, as the logged rating score is an important indicator that reflects the satisfaction of the consumers who had already purchased the item; in our future work, we plan to combine rating and reviews for better risk estimation.

Intuitively, we can directly use an item's rating distribution as the risk distribution in our model.
However, there exist some problems if we do so. 
For example, some items' rating distribution may contain 0 probabilities, which means they get no feedback on certain rating scores. 
This could happen because the dataset can still be seen as a sampling from the ground truth, and it does not contain all possible outcomes. 
Thus, except for filtering out items without enough actions, we also estimate the rating distributions for those items contain 0 probabilities to get more plausible risk distributions for them; and for those who do not, we use their raw rating distributions directly. 
Inspired by \cite{zhang2015daily}, we adopt Weibull distribution to reconstruct the continuous rating distribution from limited discrete observations, since Weibull distribution is able to fit very flexible curve shapes.
The probability density function (PDF) of the numerical rating $z$ is modeled as follows:
\begin{equation}
    f(z) = \mu \eta z^{\mu - 1}\exp(-\eta z^{\mu})
\end{equation}
where $\eta$ is the scale parameter and $\mu$ is the shape parameter.
Since the rating is a discrete value representing an approximation of user satisfaction, we use the interval integration of the estimated Weibull distribution as the probabilities of the potential outcomes.
Specifically, $p_i = \Pr(z = i) = \int_{i-0.5}^{i+0.5} f(z) \text{d}z$ is the probability that the rating score of the item is $i$ ($i$ = 1, 2, 3, 4 or 5, for $i = 1$, integral lower bound is $-\infty$ and for $i=5$, upper bound is $+\infty$).

\subsection{Risk-aware Recommendation (RARE)}
In this work, we propose an optimization framework based on discrete choice model for our \textbf{r}isk-\textbf{a}ware \textbf{re}commendation learning framework --- \textbf{RARE}.
Online shopping requires consumers to choose their desired items among alternatives, which can be considered as a discrete choice problem. 
This discrete choice problem describes a situation when a consumer chooses an option between two or even more discrete alternatives. 
More formally, consumer $u$ chooses item $v$ over a set of some other alternative products $\Omega_{u}({v})$. 
We define the total choice set as $\Pi_{u} = \{v, \Omega_{u}({v})\}$ and its $k$-th element is $\Pi^k_{u}$ (i.e., $\Pi^1_{u} = v$). 
The probability that consumer $u$ chooses alternative $v$ is denoted as $P_{uv}$.
We can simply view $v$ and $\Omega_{u}({v})$ as positive and negative training records, respectively.

Researchers in economics have utilized Random Utility Models (RUM) to deal with discrete choice problem \cite{domencich1975urban}. Different from traditional RUMs, we adopt the idea of choosing the alternative item that provides the highest utility in choosing the alternative item with the highest prospect value.
In this way, we have:
\begin{equation}
\hat{\mathcal{V}}_{u}(\Pi^k_{u}) = \mathcal{V}_{u}(\Pi^k_{u}) + \epsilon_k
\end{equation}
where $\mathcal{V}_{u}(\Pi^k_{u})$ represents the true prospect value of the $k$-th product in item set $\Pi_{u}$, and $\hat{\mathcal{V}}_{u}(\Pi^k_{u})$ represents the observed prospect value of that product and $\epsilon_k$ is a random variable capturing the impact of all unknown factors. Thus, the probability of a customer choosing $\Pi^1_{u}$ (which is positive item $v$) over other alternatives is:
\begin{equation}
\small{
P_{uv}\left(\hat{\mathcal{V}}_{u}(\Pi^1_{u}) > \hat{\mathcal{V}}_{u}(\Pi^k_{u}) \right)
= P_{uv}\left(\epsilon_k - \epsilon_1 < \mathcal{V}_{u}(\Pi^1_{u}) - \mathcal{V}_{u}(\Pi^k_{u})\right)
}
\end{equation}
where $k = 2$, ... , $|\Pi_{u}|$. If $\epsilon_1$ and $\epsilon_k$  follow an i.i.d. extreme value distribution, it can be shown that the probability of choosing $\Pi_{u}^1$ is the following multinomial logistic (MNL) model \cite{hausman1984specification},
\begin{equation}\label{eq:multilogmodel}
\small{
P(y_{uv}=1)  =   P_{uv}\left(\hat{\mathcal{V}}_{u}(\Pi^1_{u}) > \hat{\mathcal{V}}_{u}(\Pi^k_{u}) \right)
             =   \frac{\exp(\mathcal{V}_u(\Pi^1_{u}))}{\sum_{k=1}^{|\Pi_{u}|}\exp(\mathcal{V}_{u}(\Pi^k_{u}))}
             }
\end{equation}
where $y_{uv}$ is an indication function that is
\begin{equation}
    y_{uv}=\left\{
    \begin{array}{cc}
         1 & \mathcal{V}_{u}(\Pi^1_{u})>\mathcal{V}_{u}(\Pi^k_{u}) \quad\forall k \neq 1\\
         0 & \text{Otherwise}
    \end{array}.
    \right.
\end{equation}
$y_{uv}=1$ indicates user $u$ chooses item $v$ among all the other alternatives $\Omega_u(v)$.
Therefore, $P(y_{uv}=1)$ tends to maximize the probability of choosing $v$ among $\Pi_u$ as $\Pi_u^1=v$, further more, it tends to maximize the probability that the prospect value of $v$ is greater than those of its alternatives, namely, $\mathcal{V}_{u}(v)>\mathcal{V}_{u}(\Pi^k_{u})$, $\forall k \neq 1$.
In the training process, we set the number of alternatives as two and pick them randomly from items that the user has no interaction before; thus, we get $\Pi_{u} = \{v, negative\_sample_{1}, negative\_sample_{2}\}$ for each user-item pair in our train set.
 
Given the observed transactions and MNL, the model parameters can be learned by maximizing the probability of choosing the purchased items, which is
\begin{equation}
    \max \sum_{(u,v)\in\mathcal{R}}\log \left(P(y_{uv}=1)\right)-\lambda\|\Phi\|^2, \label{eq:upper}
\end{equation}
where $\mathcal{R}$ is the training dataset; $\Phi$ is the parameter set to be learned in the corresponding loss function, which will be crystallized in the following section.

\subsection{Model Specification}
Economic principles mainly focus on finding out the general rules for the majority of people, however, with the help of machine learning and massive data, we now gain the ability to design a model to fit the personalized preference of each individual consumer. 

To obtain personalized measurements, each user-item pair comes with a set of parameters, i.e. $\alpha_{uv},\beta_{uv},\lambda_{uv},\gamma_{uv},\delta_{uv},\hat{r}_{u}$. In particular, each parameter (except for the reference point $\hat{r}_{u}$) can be decomposed into the global bias, user bias, item bias, and $K$-dimensional latent factors of users and items, which is the same idea implemented in SVD based CF \cite{ricci2011introduction}. For example, the parameter $\alpha$ in the value function Eq.\eqref{eq:value} can be rewritten as:
\begin{equation}
\begin{small}
\alpha_{uv}= g^\alpha + b_u^\alpha + l_v^\alpha+{\mathbf{p}_u^\alpha}^T\mathbf{q}_v^\alpha
\end{small}
\end{equation}
where $g^\alpha$ is the global bias of $\alpha$, $b_u^\alpha$ is the user bias of $\alpha$, $l_v^\alpha$ is the item bias of $\alpha$, $\mathbf{p}_u^\alpha$ and $\mathbf{q}_v^\alpha$ are $k$-dimensional latent factors of user $u$ and item $v$. 
The above reparametrization is also used for $\beta_{uv}$, $\lambda_{uv}$, $\gamma_{uv}$ and $\delta_{uv}$. 

Besides, since each consumer has his/her own reference point, the reference point $\hat{r}_u$ of each consumer $u$ also needs to be learned. In this way, the optimization problem in Eq.\eqref{eq:upper} can be specialized as follow:
\begin{equation}
\small{
    \begin{aligned}
    \max& \sum_{(u,v)\in\mathcal{R}}\log\left(\frac{\exp(\mathcal{V}_u(v))}{\sum_{k=1}^{|\Pi_{u}|}\exp(\mathcal{V}_{u}(\Pi^k_{u}))}\right)-\lambda\|\Phi\|^2,\\
    &s.t.\quad \alpha_{uv}, \beta_{uv}, \lambda_{uv}, \gamma_{uv},\delta_{uv} \in (0,1)
    \end{aligned}
    }
\end{equation}
where $\mathcal{V}_{u}(\Pi^k_{u})$ represents the prospect value of the $k$-th product in itemset $\Pi_{u}$ and can be calculated by Eq.\eqref{eq:value}, Eq.\eqref{eq:weight} and Eq.\eqref{eq:prospectfun}. 
Meanwhile, we set the reference point $\hat{r}_u$ in Eq.\eqref{eq:value} as a learnable parameter for each consumer $u$. As a result, the parameter set is
\begin{equation}
\begin{aligned}
    \Phi=\{&g^\alpha, g^\beta, g^\lambda, g^\gamma, g^\delta, \\
&\forall u: \{\hat{r}_u, b_u^\alpha,\mathbf{p}_u^\alpha,b_u^\beta,\mathbf{p}_u^\beta,b_u^\lambda,\mathbf{p}_u^\lambda,b_u^\gamma,\mathbf{p}_u^\gamma,b_u^\delta,\mathbf{p}_u^\delta\},\\
&\forall v:\{l_v^\alpha,\mathbf{q}_v^\alpha\,l_v^\beta,\mathbf{q}_v^\beta,l_v^\lambda,\mathbf{q}_v^\lambda,l_v^\gamma,\mathbf{q}_v^\gamma,l_v^\delta,\mathbf{q}_v^\delta\}\}.
\end{aligned}
\end{equation}

Suppose we have $n$ users and $m$ items in our dateset and we set the dimension of the latent embedding as $k$, for each reparameterized parameter, which can be seen as a matrix $X$ (X = $\alpha,\beta,\lambda,\gamma,\delta$) with elements represented as $x_{uv}$, we will have a one-dimension global bias vector, a $n$-dimension user bias vector, a $m$-dimension item bias vector, a $n \times k$ user embedding matrix and a $m \times k$ item embedding matrix. 
Thus, the total number of the learning parameters is $5+5\times(n+m)(k+1)$ plus $n$, represented for a $n$-dimension reference point vector $\hat{r}$ with elements represented as $\hat{r}_{u}$.
We use log softmax loss as the loss function and batch stochastic gradient descent (SGD) as an optimizer to learn the model parameters.
Meanwhile, people's risk attitudes may not be static and may vary over time. 
To overcome this, we can retrain our model based on their recent interaction with the system and update consumers' personalized risk preferences.

\subsection{Top-K Recommendation}
Once we learned the model parameters $\alpha_{uv}$, $\beta_{uv}$, $\lambda_{uv}$,$\gamma_{uv}$, $\delta_{uv}$  and $\hat{r}_u$ according to our model, we can calculate the prospect value $V_{uv}$ for each user-item pair to rank all the products for a user. Then we select the top $K$ items to generate the top-K recommendation list.

\section{Experiments}\label{sec:experiments}

\subsection{Dataset Description}
We use the consumer transaction data from different sources --- $Amazon$\footnote{http://jmcauley.ucsd.edu/data/amazon/} \cite{he2016ups,mcauley2015image}, $Movielens$ \footnote{https://grouplens.org/datasets/movielens/} \cite{Harper:2015:MDH:2866565.2827872}, $Ciao$  and $Epinions$ \footnote{https://www.cse.msu.edu/\textasciitilde tangjili/datasetcode/truststudy.htm} \cite{Tang-etal12c,Tang-etal13a} --- in our experiments to verify the recommendation performance of \textbf{RARE}\footnote{https://github.com/TobyGE/Risk-Aware-Recommnedation-Model}. 
The $Amazon$ dataset includes user transaction (user id, item id, rating, etc.) and item metadata (item id, price, related item, etc.) on twenty-four product categories lasting from May 1996 to July 2014. 
We pick the largest three categories (namely $Books$, $Movies\& TV$, and $Electronics$), which have different sizes and data sparsity, for experiments. 
The original data is huge and highly sparse, which makes it challenging to evaluate.
Therefore, we first filter out items without prices, as the value function needs item prices as input; then, similar to previous work \cite{he2017neural, kang2018self}, we filter out users and items with fewer than ten interactions.
Meanwhile, we choose $Movielens1M$ dataset, which includes one million user transactions (user id, item id, rating, timestamp, etc.).
This dataset is much denser compared with the $Amazon$ datasets, $Ciao$ and $Epinions$, where each user has at least 20 ratings.
Considering that three $Amazon$ datasets and $Movielens1M$ only contain one certain category of items, we do the experiments on another two widely used datasets --- $Ciao$ and $Epinions$, which contain items from different categories.
Especially, $Ciao$ contains items purchased by users in 28 categories, and $Epinions$ has 27 categories in total.
Since we need prices to calculate the value function, which are not recorded by $Movielens1M$, $Ciao$, and $Epinions$, we set the price of each item in these three datasets equal to the same number, for example, \$1.
For each dataset, we sort the transactions of each consumer according to the purchase timestamp and then split the records into training, validation, and testing sets chronologically.
We then adopt the widely used leave-one-out evaluation \cite{he2017neural,cheng2016wide,kang2018self}, concretely, the test set contains user's most recent transaction, the validation set contains user's second most recent transaction, and all remaining interactions are for training.
Some basic statistics of the experimental datasets are shown in Table \ref{tab:dataset}. 
We notice that $Movielens1M$ is the densest dataset and has adequate actions per user and per item, while the three $Amazon$ datasets have more users, items, and interactions compared with $Movielens1M$.

\begin{table}
\caption{Basic statistics of the experimental datasets.}
\label{tab:dataset}
\centering
\setlength{\tabcolsep}{5pt}
\begin{adjustbox}{max width=\linewidth}
\begin{tabular}
    {lcccccccc} \toprule
    Dataset & \#users & \#items & \#act./user & \#act./item & \#act. & density \\\midrule
    Movies\& TV & 25,431 & 10,470 & 28.6 & 69.4 & 726,857 & 0.273\%\\
    Electronics & 40,983 & 16,286 & 13.5 & 34.2 &556,227 & 0.083\% \\
    Books & 157,510 & 126,578 & 29.5 & 36.8 & 4653780 & 0.023\%\\
    Movielens1M & 6040 & 3706 & 166 & 270 &1,000,209 & 4.468\%\\
    Ciao & 2248 & 16861 & 16.0 & 2.14 & 36065 & 0.095\%\\
    Epinions & 22164 & 296277 & 41.6 & 3.11 & 922267 & 0.014\%\\\bottomrule
\end{tabular}
\end{adjustbox}
\end{table}

\begin{table}
\caption{Hyperparameter settings for each dataset.}
    \centering
    \begin{tabular}{l c c c}
    \toprule
        Dataset & Latent Size &  Learning Rate & L2 Penalty \\
        \midrule
        Movies\&TV & 64 & 1e-3 & 1e-1\\
        Electronics & 16 & 1e-3 & 1e-1 \\
        Books & 256 & 5e-3 & 1e-1 \\
        Movielens1M & 16 & 1e-4 & 1e-1\\
        Ciao & 16 & 1e-2 & 1e-1\\
        Epinions & 64 & 1e-2 & 1e-1\\\bottomrule
        
    \end{tabular}
    \label{tab:parameters}
    \vspace{-10pt}
\end{table}

\begin{table}[]
\caption{Summary of the performance on six datasets. We evaluate for ranking ($F_1$ and $NDCG$, in percentage (\%) values), and $K$ is the length of recommendation list. When RARE is the best, its improvements against the best baseline are significant at p < 0.01.}
\centering
\begin{adjustbox}{max width=\linewidth}
\setlength{\tabcolsep}{7pt}
\begin{tabular}
    {m{1.3cm} l ccc ccc} \toprule
    \multirow{2}{*}{Dataset} & Metrics &  \multicolumn{3}{c}{$F_1$ Score (\%)} & \multicolumn{3}{c}{NDCG (\%)}\\\cmidrule(lr){2-2} \cmidrule(lr){3-5} \cmidrule(lr){6-8}
    & K & 5 & 10 & 20 & 5 & 10 & 20 \\\midrule 
\multirow{5}{*}{Movies\&TV} & SVD & 5.288 & 4.393 & 3.302 & 12.29 & 15.13 & 17.86 \\
&BPR-MF & 6.010 & 5.496 & 4.488 & 13.56 & 17.72 & 22.07\\
&NCF & 6.357 & 5.346 & 4.192 & \underline{15.51} & 19.17 & 23.05\\
&MPUM & \underline{6.732} & \underline{5.865} & \underline{4.573} & 15.31 & \underline{19.43} & \underline{23.77}\\
&RARE & \textbf{7.144} & \textbf{6.024} & \textbf{4.710} & \textbf{16.67} & \textbf{20.68} & \textbf{24.89}\\\midrule
\multicolumn{2}{c}{Relative Improvements} & +6.12 & +2.71 & +3.00 & +7.48 & +6.34 & +4.71\\\bottomrule

\multirow{5}{*}{Electronics} & SVD & 5.277 & 3.932 & 2.778 & 12.96 & 14.96 & 16.90 \\
&BPR-MF & 6.213 & 5.144 & 4.005 & 14.64 & 17.94 & 21.50\\
&NCF & \underline{6.815} & 5.397 & 3.968 & \underline{16.29} & \underline{19.46} & 22.55\\
&MPUM & 6.783 & \underline{5.566} & \underline{4.129} & 15.94 & 19.45 & \underline{22.74}\\
&RARE & \textbf{7.395} & \textbf{5.881} & \textbf{4.279} & \textbf{17.66} & \textbf{21.14} & \textbf{24.38}\\\midrule
\multicolumn{2}{c}{Relative Improvements} & +8.51 & +5.66 & +3.63 & +8.41 & +8.63 & +7.21\\\bottomrule

\multirow{5}{*}{Books} & SVD & 5.330 & 3.992 & 2.815 & 13.04 & 15.10 & 17.05 \\
&BPR-MF & 9.639 & 7.489 & 5.265 & 23.12 & 27.33 & 30.98\\
&NCF & \underline{10.32} & \underline{7.643} & \underline{5.282} & \underline{25.72} & \underline{29.52} & \underline{32.90}\\
&MPUM & 7.724 & 6.487 & 4.886 & 17.99 & 22.27 & 26.31\\
&RARE & \textbf{10.79} & \textbf{7.902} & \textbf{5.395} & \textbf{27.00} & \textbf{30.81} & \textbf{34.22}\\\midrule
\multicolumn{2}{c}{Relative Improvements} & +4.55 & +3.39 & +2.14 & +4.98 & +4.37 & +4.01 \\\bottomrule

\multirow{5}{*}{Movielens} & SVD & 13.02 & 9.037 & 5.763 & 32.77 & 36.44 & 39.24 \\
&BPR-MF & 14.51 & 10.44 & 6.944 & 36.19 & 40.97 & 44.97\\
&NCF & 15.17 & \underline{10.60}  & \underline{6.861}  & 39.69 & \underline{44.10}  & \underline{47.66} \\
&MPUM & \underline{15.53}  & 10.29 & 6.468 & \underline{40.67}  & 44.09 & 47.03\\
&RARE & \textbf{16.88} & \textbf{11.21} & \textbf{7.064} & \textbf{44.25} & \textbf{47.92} & \textbf{51.25}\\\midrule
\multicolumn{2}{c}{Relative Improvements} & +8.69 & +5.75 & +2.96 & +8.80 & +8.66 & +7.53 \\\bottomrule

\multirow{5}{*}{Ciao} & SVD & 7.103 & 4.812 & 3.258 & 18.54 & 20.33 & 22.32 \\
&BPR-MF & \underline{10.49} & 7.190 & \underline{4.588} & \underline{26.97} & \underline{29.75} & \underline{31.98}\\
&NCF & 7.770 & 5.670 & 4.029 & 19.63 & 22.33 & 25.19\\
&MPUM & 10.02 & \underline{7.401} & 4.550 & 23.66 & 27.33 & 29.19\\
&RARE & \textbf{11.54} & \textbf{8.250} & \textbf{5.275} & \textbf{28.23} & \textbf{31.96} & \textbf{34.56}\\\midrule
\multicolumn{2}{c}{Relative Improvements} & +10.0 & +11.5 & +15.0 & +4.67 & +7.43 & +8.07\\\bottomrule

\multirow{5}{*}{Epinions} & SVD & 11.98 & 7.604 & 4.345 & 31.69 & 33.71 & 34.76 \\
&BPR-MF & \underline{19.01} & 11.73 & 6.766 & \underline{50.79} & \underline{53.40} & \underline{55.09}\\
&NCF & 18.06 & 11.46 & 6.791 & 47.86 & 50.92 & 53.07\\
&MPUM & 18.31 & \underline{12.03} & \underline{7.136} & 46.74 & 50.65 & 52.90\\
&RARE & \textbf{19.91} & \textbf{12.78} & \textbf{7.416} & \textbf{52.31} & \textbf{55.95} & \textbf{58.11}\\\midrule
\multicolumn{2}{c}{Relative Improvements} & +4.73 & +6.23 & +3.92 & +2.99 & +4.78 & +5.48\\\bottomrule

\end{tabular}\label{tab:result}
\end{adjustbox}
\vspace{-10pt}
\end{table}

\subsection{Experimental Setup}
We compare our model with the following baselines, including both economic and non-economic methods. For economic methods, we involve baselines that do not consider risk preferences to illustrate the importance of risk consideration.

{\bf SVD}: Collaborative Filtering based on matrix factorization is a representative method for rating prediction. 
Basically, the user and item rating vectors are considered as the representation vector for each user and item.
In this experiment, we use CF based on Singular Value Decomposition techniques \cite{koren2009matrix, ricci2011introduction}.

{\bf BPR-MF}: Bayesian Personalized Ranking \cite{bpr} is one of the most widely used ranking-based methods for the top-N recommendation. 
It is considered a classification problem with two classes that bought and not bought. Instead of classifying with the prediction of a single good, BPR tried to use the difference between two predictions. 
In the implementation, we conduct balanced negative sampling on unpurchased items for model learning.

{\bf NCF}: Neural Collaborative Filtering \cite{he2017neural} is one of the state-of-the-art recommender algorithms, which is based on deep neural networks. 
In the evaluation part, we choose Neural Matrix Factorization to conduct the experiments, fusing both Generalized Matrix Factorization (GMF) and Multiple Layer Perceptron (MLP) under the NCF framework.

{\bf MPUM}: Multi-Product Utility Maximization for recommendation \cite{zhao2017multi}, which is an economic recommendation approach that maximizes the utility of product combinations for recommendation.

We implement SVD, BPR-MF, NCF and MPUM using \textit{Pytorch} \cite{paszke2019pytorch} with Adam optimizer.
For all the methods, we consider latent dimensions $d$ from \{16, 32, 64, 128, 256\}, learning rate $lr$ from \{1e-1, 5e-2, 1e-2, ..., 5e-4, 1e-4\}, and the L2 penalty is chosen from \{0.01, 0.1, 1\}. 
We tune the hyper-parameters using the validation set and terminate training when the performance on the validation set does not change too much within 20 epochs.
Meanwhile, to avoid heavy computation on all testing user-item pairs, we followed the mechanism in \cite{he2017neural, elkahky2015multi, koren2008factorization}. 
For each user $i$, we randomly sample 100 negative items (items that are not interacted by the user) and rank these items with the positive sample in the test set. 
Based on the ranking results of the 101 items, we adopt two common Top-K metrics --- $F1$ Score and $NDCG$ --- to evaluate each recommender model's performance.
$F1 Score$ considers both the precision and the recall of the test to compute the score, while $NDCG$ is a position-aware metric, involving a discount function over the rank.       
We implement \textbf{RARE} with $Pytorch$ and fine-tune hyperparameters on our validation set, and the detailed settings on each dataset for it are shown in Table \ref{tab:parameters}.

\begin{figure}[t]
\captionsetup[sub]{font=small,labelfont=normalfont,textfont=normalfont}
\centering
\hspace{-15pt}
\begin{subfigure}{0.24\textwidth}
\label{fig:ElectronicsConvRate}
        \includegraphics[scale=0.31]{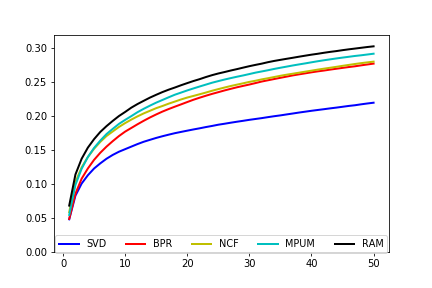}
        \caption{Movies\&TV}
\end{subfigure}
\begin{subfigure}{0.24\textwidth}
\label{fig:ElectronicsConvRate}
        \includegraphics[scale=0.31]{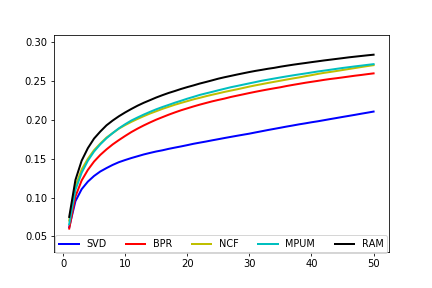}
        
        \caption{Electronics}
\end{subfigure}
\medskip
\hspace{-15pt}
\begin{subfigure}{0.24\textwidth}
\label{fig:ElectronicsConvRate}
        \includegraphics[scale=0.31]{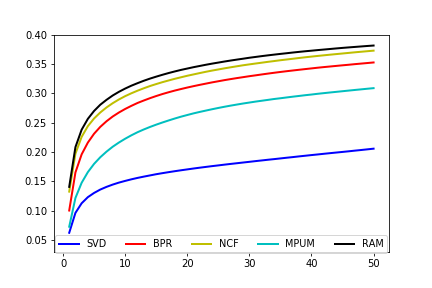}
        
        \caption{Books}
\end{subfigure}
\begin{subfigure}{0.24\textwidth}
\label{fig:ElectronicsConvRate}
        \includegraphics[scale=0.31]{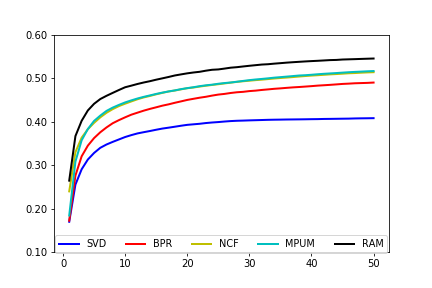}
        
        \caption{Movielens1M}
\end{subfigure}
\medskip
\hspace{-15pt}
\begin{subfigure}{0.24\textwidth}
\label{fig:ElectronicsConvRate}
        \includegraphics[scale=0.31]{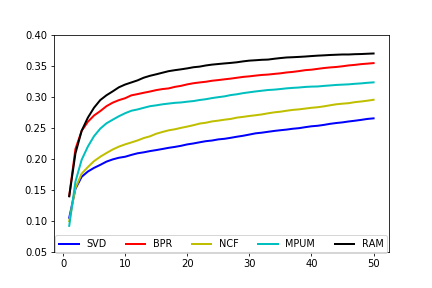}
        
        \caption{Ciao}
\end{subfigure}
\begin{subfigure}{0.24\textwidth}
\label{fig:ElectronicsConvRate}
        \includegraphics[scale=0.31]{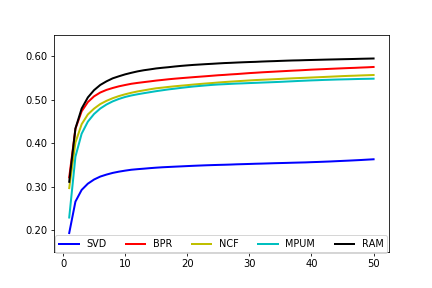}
        
        \caption{Epinion}
\end{subfigure}
\caption{NDCG on six datasets. $x$-axis is the length of the recommendation list and $y$-axis is the NDCG.}
\label{fig:NDCG}
\vspace{-15pt}
\end{figure}

\subsection{Experimental Results}
The major experimental results are shown in Table \ref{tab:result}, besides, we also plot the $NDCG$ in Figure \ref{fig:NDCG} under different length of recommendation list $K$ (from 1 to 50). We analyze and discuss the results in terms of the following perspectives.

\subsubsection*{\bf i) Recommendation Performance:} 
The results of the recommendation performance are shown in Table \ref{tab:result}.
The largest value on each dataset and for each evaluation measure is significant at 0.01 level.
Among all the baseline models, we can see that all pair-wise learning methods (BPR, NCF, MPUM, and RARE) are much better than the simple point-wise (SVD) method, which demonstrates the superiority of pair-wise learning to rank methods on top-K ranking tasks.
Further, among all the baselines, BPR is the strongest: when averaging across all datasets and recommendation lengths, BPR gets 41.0\% improvement than SVD, 2.27\% than NCF, and 1.60 \% than MPUM.

Our RARE approach achieves the best top-K recommendation performance against all baselines on six datasets from four different sources. 
On the one hand, for F1 Score, we get 10.11\% improvement than the BPR baseline when averaged across $K$ on all of the six datasets, and the largest improvement (18.87\%) is achieved when $K=5$ on \textit{Movies\&TV} dataset. 
On the other hand, for NDCG, RARE relatively improves 12.83\% than BPR on average.
Compared with deep recommendation baseline NCF, our model gets 13.82\% improvement on F1 on average, and especially a 48.52\% improvement for F1@5 on \textit{Ciao} dataset; furthermore, our model also gets 13.46\% improvement than NCF for NDCG on average, and a 43.81\% improvement for NDCG@5 on \textit{Ciao} dataset.
Meanwhile, in order to prove the validity of our improvement, we also plot the NDCG for different lengths of recommendation lists on all datasets, shown in Fig.\ref{fig:NDCG}. 
We can find that RARE outperforms all baseline for larger $K$, for example, $K=50$. 
As shown, the black curve represented as RARE, is always above the best baselines in all six datasets.
These observations imply that by modeling user behaviors under uncertainty based on established risk-aware principles, our model has the ability to capture better user preferences resulting in better recommendation results.

\subsubsection*{\bf ii) Behavioral Economics vs. Classical Economics:} 
We also have a strong economic recommender baseline --- MPUM.
When comparing with MPUM, our model gets 12.97\% improvement on average.
Especially, for the F1 Score, RARE gets 10.45\% improvement against MPUM; for NDCG, the improvement is larger, increasing to 15.48 \%.
We believe there are several reasons to explain this improvement.
First, the economic intuition behind MPUM is to make use of the substitutability and complementarity between products. 
However, most of our datasets only contain items from one category, which may decrease their performance. 
Second, the utility function implemented in their model is the well-known Constant Elasticity Substitution (CES) utility function, which is always risk-aversion \cite{10.1007/978-3-319-13359-1_24}, while Prospect Theory Value Function is a more delicate and comprehensive method that involves the risk attitudes and considers the difference between gains and losses, which allows for finer-grained modeling of consumer behaviors in online commercial environments.

\begin{table}
\caption{Summary of the performance against ablations on six datasets. When RARE is the best, its improvements against the best ablation are significant at p < 0.01.}
\begin{adjustbox}{max width=\linewidth}
\setlength{\tabcolsep}{7pt}
\begin{tabular}
    { m{1.3cm} l ccc ccc } \midrule
    \multirow{2}{*}{Dataset} & Metrics &  \multicolumn{3}{c}{$F_1$ Score (\%)} & \multicolumn{3}{c}{NDCG (\%)}\\\cmidrule(lr){2-2} \cmidrule(lr){3-5} \cmidrule(lr){6-8}
    & K & 5 & 10 & 20 & 5 & 10 & 20 \\\midrule 
\multirow{4}{*}{Movies\&TV} & RARE-WF & 4.749 & 3.772 & 2.702 & 11.11 & 13.34 & 15.31 \\
&RARE-VF & 2.015 & 1.961 & 2.013 & 4.666 & 6.274 & 8.919 \\
&RARE-RP & 3.872 & 3.100 & 2.191 & 9.294 & 11.15 & 12.69 \\
&RARE & \textbf{7.144} & \textbf{6.024} & \textbf{4.710} & \textbf{16.67} & \textbf{20.68} & \textbf{24.89}\\\midrule

\multirow{4}{*}{Electronics} & RARE-WF & 6.585 & 5.418 & 3.182 & 16.28 & 18.96 & 22.05 \\
&RARE-VF & 1.805 &2.082 & 2.126 & 3.784 & 5.830 & 8.619 \\
&RARE-RP & 4.387 & 3.118 & 2.101 & 10.96 & 12.37 & 13.64 \\
&RARE & \textbf{7.395} & \textbf{5.881} & \textbf{4.279} & \textbf{17.66} & \textbf{21.14} & \textbf{24.38}\\\midrule

\multirow{4}{*}{Books} & RARE-WF & 4.979 & 3.620 & 2.450 & 12.48 & 14.19 & 15.70 \\
&RARE-VF & 2.116 & 2.322 & 2.405 & 4.508 & 6.687 & 9.880 \\
&RARE-RP & 4.670 & 3.392 & 2.292 & 11.68 & 13.28 & 14.68 \\
&RARE & \textbf{10.79} & \textbf{7.902} & \textbf{5.395} & \textbf{27.00} & \textbf{30.81} & \textbf{34.22}\\\midrule

\multirow{4}{*}{Movielens} & RARE-WF & 14.42 & 8.823 & 5.065 & 39.28 & 41.11 & 42.33 \\
&RARE-VF & 2.130 & 1.860 & 1.788 & 5.320 & 6.610 & 8.786 \\
&RARE-RP & 8.742 & 5.205 & 4.092 & 24.75 & 25.55 & 29.16 \\
&RARE & \textbf{16.88} & \textbf{11.21} & \textbf{7.064} & \textbf{44.25} & \textbf{47.92} & \textbf{51.25}\\\midrule

\multirow{4}{*}{Ciao} & RARE-WF & 9.356 & 6.584 & 4.135 & 23.51 & 26.34 & 28.22 \\
&RARE-VF & 4.864 & 3.680 & 2.453 & 11.53 & 13.49 & 14.92 \\
&RARE-RP & 9.060 & 6.446 & 4.007 & 22.55 & 25.41 & 27.11 \\
&RARE & \textbf{11.54} & \textbf{8.250} & \textbf{5.275} & \textbf{28.23} & \textbf{31.96} & \textbf{34.56}\\\midrule

\multirow{4}{*}{Epinions} & RARE-WF & 9.475 & 5.866 & 3.370 & 25.46 & 26.79 & 27.61 \\
&RARE-VF & 8.727 & 7.501 & 4.589 & 20.14 & 25.32 & 27.19  \\
&RARE-RP & 8.484 & 5.284 & 3.153 & 14.79 & 24.12 & 25.16  \\
&RARE & \textbf{19.91} & \textbf{12.78} & \textbf{7.416} & \textbf{52.31} & \textbf{55.95} & \textbf{58.11}\\\midrule

\end{tabular}\label{tab:ablation_result}
\end{adjustbox}
\vspace{-10pt}
\end{table}

\subsection{Ablation Studies}

In order to have a better understanding of the role of each part of our model, i.e., the value function and weighting function, we further conduct ablation studies to analyze our model 
and answer the following research questions:
\begin{itemize}
    \item Whether users have their own judgments of the extent of gains and losses?
    \item Whether users have their own sensitivity of the happening of possible events?
    \item Whether users have their own expectations of the outcome of probabilistic events?
\end{itemize}

Based on the above questions, we design three variants of RAM and compare RAM with each of them. The detailed forms of these variants are shown as following.

{\bf RARE without Value Function (RARE-VF)}: Prospect theory for recommendation without using the personalized value function, is a variant of our model that does not consider the personalized value judgments introduced by value function. 
We keep the global bias of each parameter in the original value function (e.g. $g_{\alpha}$, $g_{\beta}$, $g_{\lambda}$) and remove the remaining personalized parts to create the non-personalized one, shown as Eq. \eqref{eq:vf}; meanwhile, the weighting function in this variant is still be the same as Eq. \eqref{eq:weight}.

\begin{equation} \label{eq:vf}
    v(x_{i})=
    \begin{cases}
    g_{\lambda}x_{i}^{g_{\alpha}} & \text{if } x_{i}\ge0\\
    -(-x_{i})^{g_{\beta}}& \text{if } x_{i}<0\\
    \end{cases}\\
\end{equation}



{\bf RARE without Weighting Function (RARE-WF)}: Prospect theory for recommendation without using the personalized weighting function is another variant of our model, which does not consider the personalized nonlinear transformation introduced by weighting function. 
Instead, it uses the original probabilities directly, which is shown in Eq. \eqref{eq:wf}; meanwhile, the value function of this variant is still the same as Eq. \eqref{eq:value}.
\begin{equation} \label{eq:wf}
    \pi(p_{i})= p_{i} 
\end{equation}

{\bf RARE without Reference Point (RARE-RP)}: Prospect theory for recommendation without using the reference point, which is the third variant of our model.
A very fundamental idea in prospect theory is that people have different risk attitudes towards gains and loss, by removing reference point, we eliminate this cornerstone from the theory.
Since we do not use reference point $\hat{r}$ in this case, there will only exist gains for consumers, because $x_{i} = price\cdot\tanh(r_{i}) \geq 0$.
The (positive) value function and (positive) weighting function used in this variant are shown as follows,

\begin{equation}
    v(x_{i})=x_{i}^\alpha
\end{equation}
\begin{equation}
    \pi(p_{i})=\frac{p_{i}^\gamma}{(p_{i}^\gamma+(1-p_{i})^\gamma)^{1/\gamma}}
\end{equation}

Key experimental results of the ablation study are shown in Table \ref{tab:ablation_result}. 
We compare, analyze, and discuss the three variants of RARE in terms of the following perspectives.

\subsubsection{\bf The role of Personalized Value Function}
In RARE-VF (RARE w/o Personalized Value Function), we set $\alpha = g_{\alpha}$, $\lambda = g_{\lambda}$ and $\beta = g_{\beta}$ in the value function and maintain the weighting function same as RARE.
From Table \ref{tab:ablation_result}, comparing with our RARE, RARE-VF decreases 66.55 \% when averaged across K on all datasets.
We can also notice that among all six datasets, RARE-VF is always the worst, which, in turn, indicates the importance of the personalized value function.
Thus, we can conclude that people do have their own subjective judgments over the objective values of the products, particularly, gains and losses.

\subsubsection{\bf The role of Personalized Weighting Function}
When we use the original probability without the weighting function, the whole method is noted as RARE-WF (RARE w/o Personalized Weighting Function). 
According to Table \ref{tab:ablation_result}, our RARE model has a better performance against RARE-WF. 
When averaged across all datasets and recommendation lengths, RARE-WF has a decline of 33.70 \% on F1 Score and 30.46 \% on NDCG.
The decline indicates that users with different risk attitudes should have different estimations, even with the same event probabilities. 
This personalized weighting function, as a nonlinear transformation of the probability scale, will predict a subjective probability for each user, so it can help the model better estimate users' risk attitudes in order to get better recommendation performances.

\subsubsection{\bf The role of Personailzed Reference Point}
The reference point is an essential part of Prospect Theory. 
It implies the consumer's subjective threshold of gain and loss.
In the third variance of RARE, i.e., RAW-RP, we set the reference point equal to 0, which means that every outcome will be gain for consumers.
In Table \ref{tab:ablation_result}, we can see that RARE achieves higher scores than RARE-RP.
When averaged across all datasets and recommendation lengths, RARE-RP relatively decreases more than 46\% on average.
In fact, setting all reference points to zero is equal to eliminate the loss of wealth; in other words, all the items are "seemingly" safe to the users.
This might let the model recommend risky items to risk-aversion people, which causes bad performances.
Thus, consumers possess a personalized threshold of gains and losses when facing different outcomes. 


\section{Conclusions and Future Work}\label{sec:conclusions}
In this paper, we introduce risk attitudes into personalized recommendation systems and propose risk-aware recommendation.
In particular, we bridge prospect theory and machine learning algorithms together to predict individual risk attitudes. 
We believe that online shopping could be risky with uncertainty, and the risk attitudes of different consumers may affect their decision making processes under uncertainty/risk. 
Understanding the risk attitude of consumers can help the system to predict human behaviors accurately for better services. 
Meanwhile, we advance prospect theory into a personalized version based on machine learning over large-scale consumer transaction logs. 
Experimental results verified the effectiveness of our model in terms of top-K recommendation. 
In the future, we will consider user risk attitudes in other online systems beyond e-commerce recommendation, and consider other economic principles and/or learning methods to benefit recommendation systems both effectively and economically.

\section*{Acknowledgement}
The authors thank the anonymous reviewers for the careful reviews and constructive suggestions.

\newpage
\bibliographystyle{ACM-Reference-Format}
\balance
\bibliography{base}

\end{document}